\documentclass{aa}
\usepackage[varg]{txfonts}
\usepackage{graphicx}
\usepackage{txfonts}
\usepackage{natbib}
\usepackage{verbatim}
\usepackage[pdftex]{color}
\usepackage[dvipsnames]{xcolor}
\usepackage{hyperref}
\usepackage{ulem}
\usepackage{subfigure}
\hypersetup{
    colorlinks = true,
    citecolor ={blue}
}
\usepackage{mathrsfs}
\bibpunct{(}{)}{;}{a}{}{,} 
\def\msun{M$_\odot$}

\def\rsun{R$_\odot$}

\usepackage{booktabs}
\definecolor{ultramarine}{rgb}{0.07, 0.1, 0.6} 
\definecolor{myblue}{rgb}{0.07, 0.2, 0.6} 
\definecolor{dopal}{rgb}{.70, .25, .05}

\begin{document}

\title{J0526+5934: a peculiar ultra-short period  double white dwarf 
}

\author{Alberto Rebassa-Mansergas\inst{1,2}\thanks{E-mail: alberto.rebassa@upc.edu},
Mark Hollands \inst{3}, Steven G. Parsons\inst{3}, Leandro G. Althaus\inst{4,5}, Ingrid Pelisoli\inst{6}, Puji Irawati\inst{7}, Roberto Raddi\inst{1}, Maria E. Camisassa\inst{1}, Santiago Torres\inst{1,2}}
\institute{Departament de F\'\i sica, 
           Universitat Polit\`ecnica de Catalunya, 
           c/Esteve Terrades 5, 
           08860 Castelldefels, 
           Spain
           \and
            Institute for Space Studies of Catalonia, 
           c/Gran Capit\`a 2--4, 
           Edif. Nexus 104, 
           08034 Barcelona, 
           Spain
           \and
           Department of Physics and Astronomy, University of Sheffield, Sheffield, S3 7RH, UK
           \and
           Facultad de Ciencias Astron\'omicas y Geof\'isicas, Universidad Nacional de La Plata, Paseo del Bosque s/n, 1900 La Plata, Argentina
           \and
           Instituto de Astrof\'isica La Plata, UNLP-CONICET, Paseo del Bosque s/n, 1900 La Plata, Argentina
           \and
           Department of Physics, University of Warwick, Coventry, CV4 7AL, UK
           \and    
           National Astronomical Research Institute of Thailand, 260 Moo 4, T. Donkaew, A. Maerim, Chiangmai, 50180 Thailand
           }

\date{Received ; accepted }
\abstract{
     Ultra-short  period  compact  binaries  are  important  sources  of
   gravitational waves, which include e.g.  the progenitors of type Ia
   supernovae or the  progenitors of merger episodes that  may lead to
   massive  and  magnetic  single   white  dwarfs.  J0526+5934  is  an
   unresolved  compact binary  star  with an  orbital  period of  20.5
   minutes that belongs to this category.  }
{
  The visible component of J0526+5934  has been recently claimed to be
  a hot sub-dwarf star with a CO  white dwarf companion. Our aim is to
  provide  strong observational  plus  theoretical  evidence that  the
  primary star is  rather an extremely-low mass  white dwarf, although
  the hot subdwarf nature cannot be completely ruled out.
}
{
  We analyse  optical spectra together with  time-series photometry of
  the visible  component of  J0526+5934 to  constrain its  orbital and
  stellar  parameters.  We  also  employ  evolutionary  sequences  for
  low-mass white  dwarfs to derive  independent values of  the primary
  mass.
}
{
  From the analysis of our observational  data, we find a stellar mass
  for the  primary star  in J0526+5934  of 0.26$\pm$0.05\,M$_{\odot}$,
  which perfectly matches the 0.237$\pm$0.035\,M$_{\odot}$ independent
  measurement   we   derived   from   the   theoretical   evolutionary
  models.  This value  is  considerably lower  than the  theoretically
  expected and  generally observed mass  range of hot  subdwarf stars,
  but falls  well within the  mass limit values of  extremely low-mass
  white dwarfs.
}
{
  We  conclude J0526+5934  is  the fifth  ultra-short period  detached
  double white dwarf currently known.
}

\keywords{(Stars:) white dwarfs; (Stars:) binaries (including multiple): close}
\titlerunning{An ultra-short period double white dwarf}
\authorrunning{Rebassa-Mansergas et al.}

\maketitle

\section{Introduction}
\label{introduction}

Together with low-mass main sequence  stars, white dwarfs are the most
common objects in our Galaxy. Indeed, over 95\% of main sequence stars
will   become,   or   have   already   turned   into,   white   dwarfs
\citep{Althaus2010}. That  is, after  all nuclear  evolutionary phases
take  place,  only the  hot  Earth-sized  core  of the  star  remains,
typically with a mass of $\simeq$0.6\,M$_{\odot}$ \citep{Hollands2018,
  Kilic2020,  McCleery2020, Jimenez2023,  Obrien2024}.  Once they  are
formed, these compact objects cool down  over periods of time that are
larger than the Hubble time, a cooling process that is relatively well
understood \citep{Blouin2019, Bauer2020, Camisassa2016, Camisassa2019,
  Camisassa2023}.  As a  consequence, white  dwarfs are  very valuable
tools that can be used  as cosmochronometers to e.g. place constraints
of     the     ages     of      open     and     globular     clusters
\citep[e.g.][]{Garcia-Berro2010, Jeffery2011,  Torres2015} as  well as
the   age  of   the  Galactic   disk  \citep[e.g.][]{Garcia-Berro1988,
  Oswalt1996} and  halo \citep[e.g.][]{Kilic2020, Torres2021}.  If the
white  dwarfs are  in wide-enough  binary systems  with main  sequence
companions (where no  mass transfer episodes took place  in the past),
they  can  be used  to  provide  ages  for  their companions  to  thus
constrain  e.g.  the   age-metallicity  relation  \citep{Rebassa2016b,
  rebassa-mansergasetal21}  and the  age-velocity dispersion  relation
\citep{Raddi2022}  of  the  Milky Way,  or  the  age-activity-rotation
relation       of       low-mass       main       sequence       stars
\citep{rebassa-mansergasetal13-1,     morganetal12-1,     Skinner2017,
  Rebassa2023}.

White dwarf  binaries are  also of extreme  interest when  the orbital
separations are short,  of the order of  a few days or  less. In these
cases, the systems most likely formed via at least one common envelope
evolution \citep{Webbink2008}. For example, post-common envelope white
dwarf-main sequence binaries  have been proven to be  very valuable to
constrain    current   theories    of   common    envelope   evolution
\citep{Camacho2014, Cojocaru2017, Zorotovic2022}  and magnetic braking
\citep{Schreiber2010,  Zorotovic2016}, as  well  as  to constrain  the
mass-radius relation of white dwarfs \citep{Parsons2017}, brown dwarfs
\citep{vanRoestel2021}, sub-dwarf stars  \citep{Rebassa2019b} and main
sequence stars \citep{Parsons2018}  if they are eclipsing.  One of the
possible  products of  post-common envelope  systems are  double white
dwarfs, considering that the main sequence companions have had time to
evolve out of  the main sequence, thus forcing the  systems to enter a
second common envelope phase. However, it  is not clear yet whether or
not double white  dwarfs form through two episodes  of common envelope
\citep{Nelemans2005, Sluys2006}, and an  alternative scenario has been
proposed that  involves a first  phase of stable  but non-conservative
mass transfer followed by  a common envelope episode \citep{Woods2012,
  Ge2015,  Schreiber2022}.  In  any case,  short-period  double  white
dwarfs are  important gravitational  wave sources,  which will  be the
dominant  sources  for  the  forthcoming  Laser  Interferometer  Space
Antenna  (LISA) mission  \citep{Korol2018}  and are  also  one of  the
favoured systems to produce type Ia supernovae \citep{Whelan+Iben1973,
  Iben+Tutukov1984,  Liu2018}.   Finding  potential  double-degenerate
supernovae  Ia  progenitors  is   extremely  challenging  via  optical
spectroscopy,  however they  are expected  to be  found through  their
detection of gravitational waves \citep{Rebassa2019a}.

\begin{figure}
   \centering
   \includegraphics[width=\columnwidth]{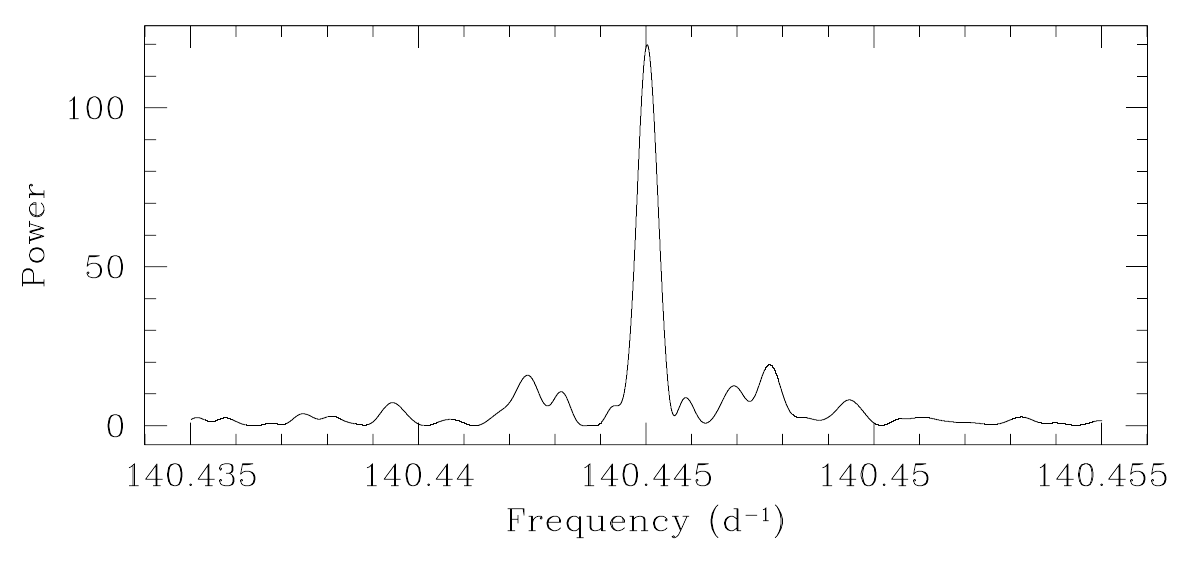}
   \caption{Scargle  periodogram  of  J0526+5934  resulting  from  our
     analysis of  the ZTF  data. The strongest  peak corresponds  to a
     period  of $\simeq$10  minutes, which  is half  the true  orbital
     period.}
\label{f-ztf}
\end{figure}

Currently,  there  are  several  hundred  double  white  dwarfs  known
\citep{Rebassa2017,  Breedt2017,  Maoz2018, Napiwotzki2020},  many  of
which  are  eclipsing \citep{Hallakoun2016,  Parsons2020,  Keller2022,
  Kosakowski2022, Munday2023}.  Of particular interest are  those that
contain     an     extremely     low-mass    (ELM)     white     dwarf
($\la$0.3\,M$_{\odot}$). These  objects cannot  be formed  in isolated
evolution,  and   are  believed  to   be  form   as  a  result   of  a
common-envelope  phase  or  after  an  episode  of  stable  Roche-lobe
overflow  \citep{Istrate2016,  Li2019}.  Most  of  the  known  objects
belonging to  this type have been  identified thanks to the  mining of
the    Sloan   Digital    Sky   Survey    spectroscopic   data    base
\citep{Gianninnas2015,  Bell2017,  Brown2020,   Brown2022}.  With  the
advent of  the astrometric and  photometric data provided by  the {\it
  Gaia} satellite  \citep{Gaia2018, Gaidos2023},  many more  ELM white
dwarfs    are    being   identified    \citep{Inight2021,    Wang2022,
  Kosakowski2023_1}. We are  currently in the process  of building and
characterising a volume-limited sample of  ELM white dwarfs using {\it
  Gaia}  plus follow-up  spectroscopy \citep{Pelisoli+Voss2019}.  As a
result of  this endeavour, we  have identified a  peculiar ultra-short
period ($<$25 minutes) double white dwarf presumably containing an ELM
white dwarf, which we present and analyse in detail in this work.

The structure of the paper is as follows. In Section\,\ref{s-j0526} we
introduce J0526+5934, the target  of study. In Section\,\ref{s-obs} we
describe our extensive follow-up  campaign. The results, including our
spectral    and    light-curve     analysis,    are    presented    in
Section\,\ref{s-results}.     We     discuss    our     results     in
Section\,\ref{s-discuss}  and  summarise  and  conclude  the  work  in
Section\,\ref{conclusions}.

\section{J0526+5934}
\label{s-j0526}

\citet{Pelisoli+Voss2019} identified J0526+5934 (RA = 81.54342º, DEC =
59.57926º; {\it  Gaia} DR3  ID = 282679289838317184)  as an  ELM white
dwarf  candidate.  Zwicky  Transient Facility,  ZTF  \citep{Masci2019,
  Dekani2020},  photometry  is  available  for  this  object  and  our
analysis  of  the  light-curve  indicated   a  very  short  period  of
$\simeq$10.25 minutes  (see Figure\,\ref{f-ztf}).  It is  worth noting
that the  periodogram did  not reveal any  significant peak  at double
this value,  which is, as  we will see in  the next section,  the true
orbital period \citep{Ren2023}.

\begin{figure}
   \centering
   \includegraphics[width=\columnwidth]{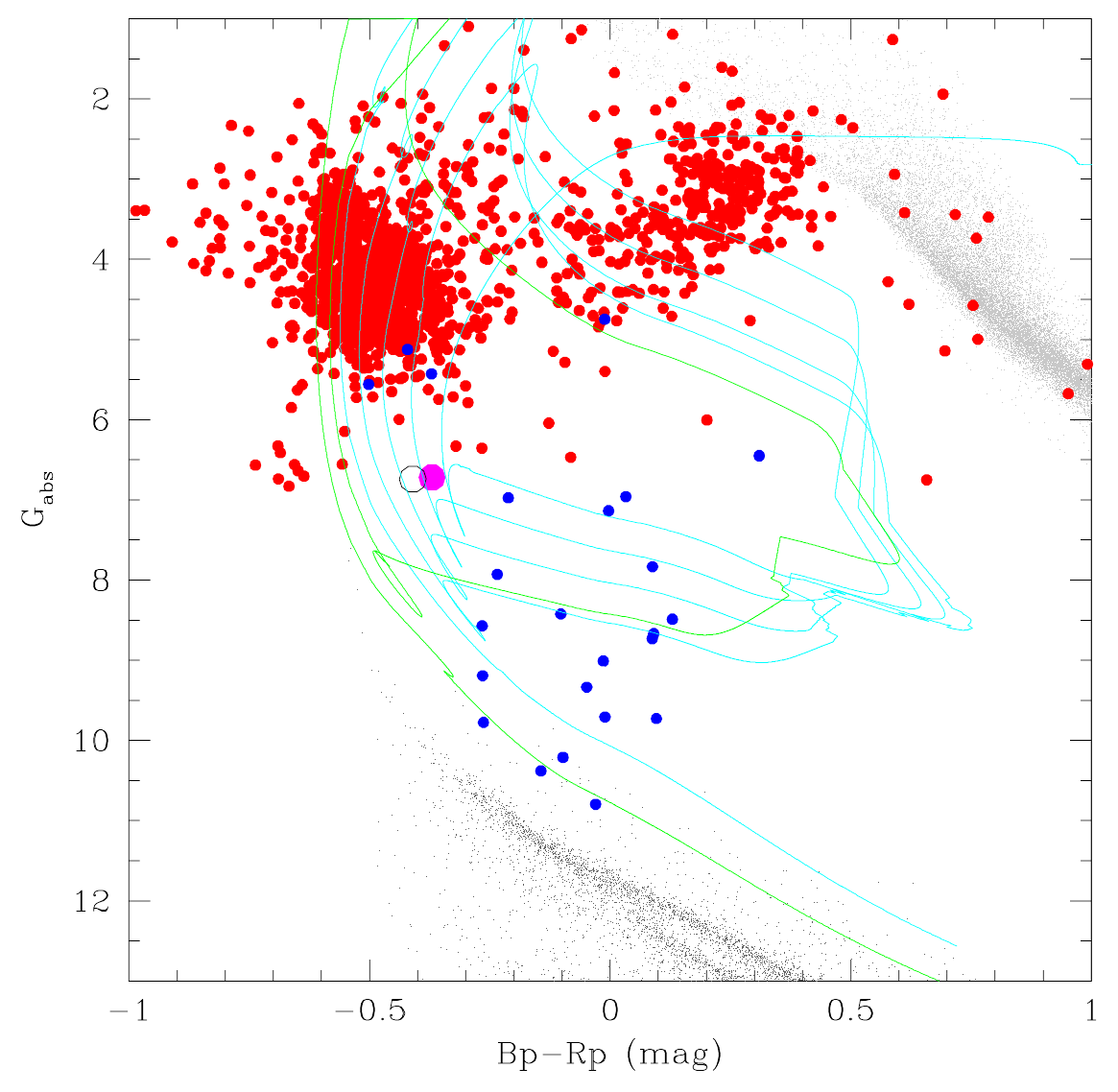}
   \caption{Absolute magnitude-colour  diagram of single  white dwarfs
     (dark grey)  and main sequence  stars (light grey) within  100 pc
     from  \citet{Jimenez2023}, hot  subdwarf  stars  (red dots)  from
     \citet{Geier2020}  and   ELM  white   dwarfs  (blue   dots)  from
     \citet{Brown2016,  Kilic2017, Pelisoli2018}.  J0526+5934 (magenta
     solid dot)  falls in the  transition region between  hot subdwarf
     stars and ELM white dwarfs.  The black open circle represents the
     G$_\mathrm{abs}$ vs. Bp-Rp that we obtain for J0526+5934 from our
     spectral  modeling (Section\,\ref{s-specfit}),  which is  in good
     agreement  with  the $Gaia$  observed  value  (magenta dot).  All
     magnitudes have been dereddened  using the extinction provided by
     the 3D  maps of  \citet{Lallement2014} except for  J0526+5934 (in
     this  case we  used  the reddening  obtained  from our  analysis;
     Section\,\ref{s-specfit})   and   only  objects   with   positive
     parallaxes   and   relative   errors   below   10\%   have   been
     considered. The  cyan and green lines  represent the evolutionary
     tracks  for a  0.226\,M$_{\odot}$  and  a 0.324\,M$_{\odot}$  ELM
     white dwarf, respectively, from \cite{Istrate2016}.}
\label{f-HR}
\end{figure}

The location of  J0526+5934 in the $Gaia$ colour  magnitude diagram is
far away from the typical white  dwarf locus, and within the region of
hydrogen   shell   flashes   of    proto   ELM   white   dwarfs   (see
Figure\,\ref{f-HR}).  As it can be seen from the figure, J0526+5934 is
also  relatively close  to the  locus occupied  by hot  subdwarfs, and
indeed it has  been claimed that the visible component  of this binary
belongs to  such category by \citet{Kosakowski2023,  Lin2023}. This is
mainly  due  to   the  fact  that  the  authors  measure   a  mass  of
$\simeq$0.36--0.38\,M$_{\odot}$  for  this   compact  object.  In  our
analysis, as  we will show  in the  forthcoming sections, we  derive a
mass for the visible  component of $\simeq$0.26\,M$_{\odot}$, which is
fully and more compatible with the  hypothesis that it is an ELM white
dwarf. Although  we find no robust  evidence to rule out  the subdwarf
nature, we argue that the visible  component of J0526+5934 is hence an
ELM white dwarf and we treat it  as such throughout the paper. We also
assume that the unseen companion  is another white dwarf (although the
possibility exists that it belongs to a more exotic category such as a
neutron star).  Under these assumptions, J0526+5934  becomes the fifth
ultra-short period detached double white dwarf known to date, together
with  SDSS\,J065133.338+284423.37  \citep[][12.7  minutes]{Brown2011},
ZTF\,J1539+5027  \citep[][6.9   minutes]{Burdge2019},  PTF\,J0533+0209
\citep[][20.6  minutes]{Burdge2019a} and  ZTF\,J2243+5242 \citep[][8.8
  minutes]{Burdge2020}.

\section{Observations}
\label{s-obs}

In  this section  we give  details  of the  follow-up observations  we
carried out at the Telescopi  Joan Or\'o, the Gran Telescopio Canarias
and the Thai National Telescope.

\subsection{Telescopi Joan Or\'o}
\label{s-tjo}

The orbital  period of J0526+5934  is too  short for obtaining  a high
signal-to-noise  spectrum  in  one single  exposure  without  avoiding
orbital smearing,  given that  it is quite  faint ($G$=17.5  mag). For
this reason, and with the aim  of measuring an orbital period accurate
enough to  plan the  spectroscopic observations, we  first followed-up
this object  with the 0.8m Telescopi  Joan Or\'o (TJO) located  in the
Montsec Observatory in Lleida, Spain.

We used the LAIA (Large Area Imager for Astronomy) instrument equipped
with  the 4k$\times$4k  Andor iKon  XL CCD  and the  Johnson-Cousins V
filter. The observations took place during the night of 2020/10/23 and
lasted for a total of  $\simeq$5 hours. Individual exposure times were
130 seconds,  reaching a signal-to-noise  ratio of $\simeq$50  for the
target in each  image. The readout time  of the CCD is  8 seconds. The
data were  automatically reduced by  the TJO internal  pipeline, which
also yields differential photometry for the target plus two comparison
stars.  The analysis  of the  TJO  photometry resulted  in an  orbital
period  of   616.00$\pm$0.66  seconds,  in  agreement   with  the  ZTF
measurement (Section\,\ref{s-j0526}; note that, in the same way as for
the  ZTF data,  the strongest  signal  corresponded to  half the  true
period).

\subsection{Gran Telescopio Canarias}
\label{s-GTC}

We  obtained  follow-up  spectroscopy  of  J0526+5934  with  the  Gran
Telescopio Canarias, GTC, at the  Observatorio Roque de los Muchachos,
La Palma. The  telescope was equipped with the  OSIRIS instrument, the
2000B grating  and the 0.6”  slit. Thus, the spectra  acquired covered
the $\simeq$3950--5700\,\AA\, wavelength range at a resolving power of
$\simeq$2100.  The observations  took place  on 2022/11/26  and lasted
$\simeq$6.5 hours.

Given the  short orbital  period of  J0526+5934, we  avoided exposures
longer than 1  minute, otherwise the spectra would  have suffered from
orbital  smearing. We  planned the  observations under  the assumption
that the orbital  period was the one we obtained  from the analysis of
the TJO  data (Section\,\ref{s-tjo}). Thus,  we aimed at  obtaining 37
cycles of 12 short exposures of 27.93+$t_\mathrm{read}$ seconds, where
$t_\mathrm{read}$  is the  CCD readout  time (23.4  seconds, including
also     the     setup     for      the     next     exposure)     and
27.93$=(P_\mathrm{orb}/12)-t_\mathrm{read}$, where $P_\mathrm{orb}$ is
the expected  orbital period.  As  a consequence, this  strategy would
allow us to take 37 spectra at  each 1/12th of the orbit in 6.5 hours,
which we would  then combine to obtain 12 spectra  equally spread over
the  entire  orbit. The  orbital  period  uncertainty of  $\simeq$0.65
seconds (Section\,\ref{s-tjo})  implies a maximum drift  of 24 seconds
after the  6.5 hours, assuming  the orbital  period was 616  s.  Given
that the total time of 51.33 seconds  (27.93 s of exposure plus 23.4 s
of readout time) is longer than the maximum drift, this implies the 37
spectra  at each  1/12th  taken over  the 6.5  hours  can be  combined
without causing any smearing.

The observations were carried out  following the above strategy and we
reduced/calibrated    the    spectra     using    the    \verb|pamela|
\citep{Marsh1989} and \verb|MOLLY|\footnote{Developed by Tom Marsh and
  available                                                         at
  \url{http://deneb.astro.warwick.ac.uk/phsaap/software}.}    packages
respectively.  When combining  the 37  spectra at  each 1/12th  of the
orbit, we found the spectra displayed a double-lined profile of nearly
identical lines. It was then that we realised these lines were exactly
the same but shifted by the same  amount both towards the blue and the
red.  In other  words,  the  orbital period  was  twice  the value  we
measured from the TJO data,  1232.00$\pm$0.66 s, in agreement with the
period reported  by \citet{Ren2023,  Kosakowski2023, Lin2023},  and we
were sampling 24 points in the orbit rather than 12. As a consequence,
we then combined the spectra at each  1/24th of the orbit. That is, we
obtained 24 spectra of J0526+5934 equally spread over the orbit.

Note that  since the orbital  period is  double what we  expected, the
maximum   drift  drops   to  12   seconds  for   the  length   of  the
observations. As a consequence, our spectra do not suffer from orbital
smearing.

\subsection{Thai National Telescope}

More follow-up photometric  data of J0526+5934 were  obtained with the
2.4m  Thai National  Telescope in  Doi Inthanon,  Thailand, using  the
ULTRASPEC  instrument   \citep{Dhillon2014}.  The   observations  were
conducted on 2020/03/29, 2020/12/10,  and 2021/02/05 where we followed
up the target for 2x the orbit, 3x the orbit, and 8x the orbit in each
respective    night.   We    opted   to    use   the    $KG5$   filter
\citep[$u'$+$g'$+$r'$,][]{Hardy2017}  for all  of our  observations to
optimise the  signal-to-noise ratio. The  data were taken  in windowed
mode  with  box  sizes  of $\simeq$3-4  arcminutes  to  ensure  enough
comparison stars within the field-of-view.

The   data   were   then   reduced   using   the   HiPERCAM   pipeline
\citep{Dhillon2007}  to obtain  the  fluxes of  J0526+5934 and  nearby
comparison stars. The  signal-to-noise ratio from our data  is 40 with
2.1s exposure time during the first observing night. The sky was clear
in  2020/03/29  with  seeing  between 2-2.5.  We  got  signal-to-noise
$\la$30 in  the second and  the third  runs due to  weather conditions
(intermittent clouds  and seeing $\la$1.5-3). The  exposure times used
in both nights was $\simeq$2s.

\section{Results}
\label{s-results}

\begin{figure}
   \centering
   \includegraphics[width=0.68\columnwidth, angle=-90]{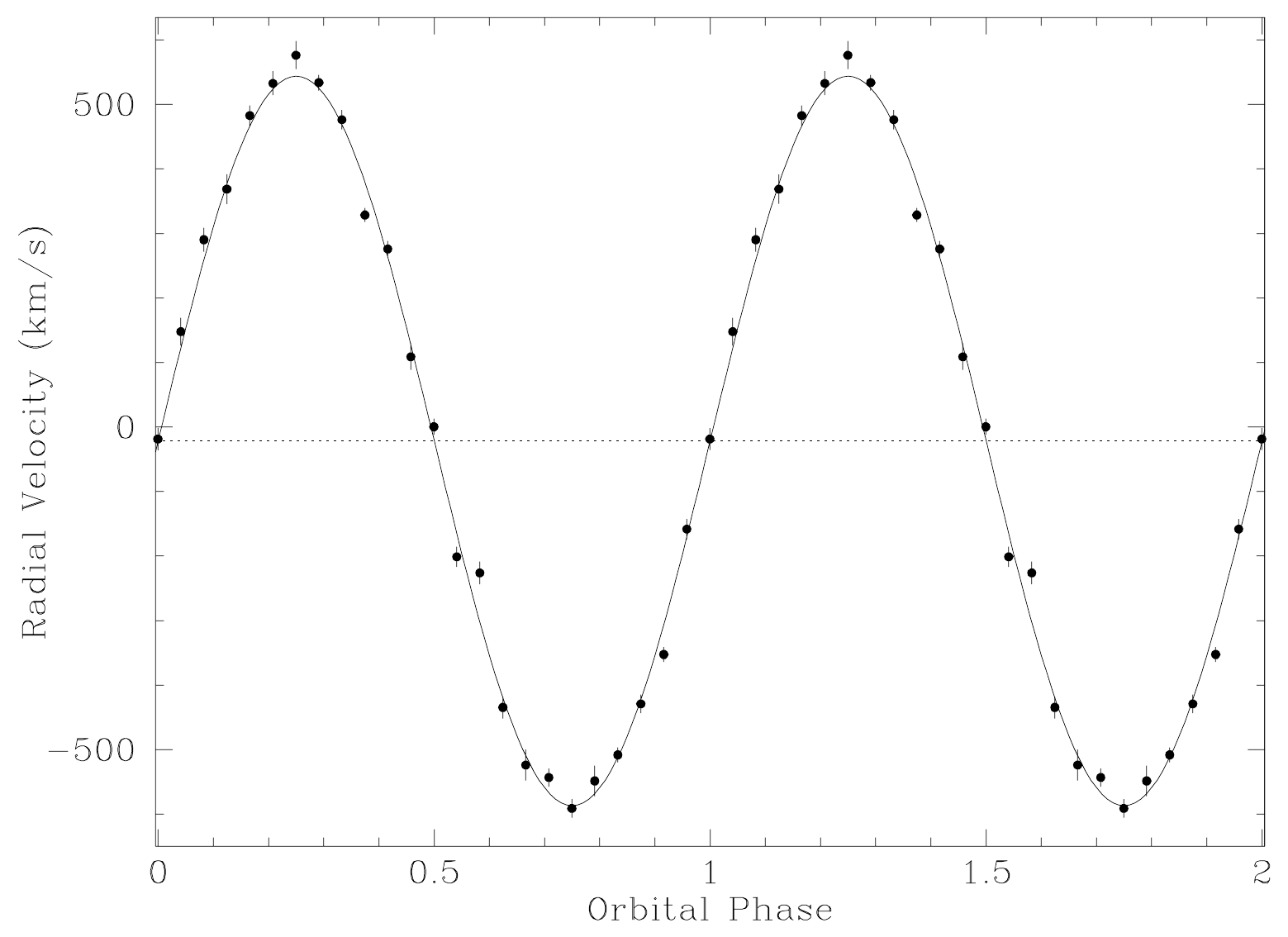}
   \caption{Radial velocities of  J0526+5934, representing the orbital
     motion  of the  ELM white  dwarf (solid  dots) as  a function  of
     orbital  phase; where  phase  0 indicates  the  time of  inferior
     conjunction. The solid line is the  best sine fit to the data and
     the horizontal dotted line indicates the systemic velocity.}
\label{f-rv_curve}
\end{figure}

\begin{figure*}
   \centering
   \includegraphics[width=\textwidth]{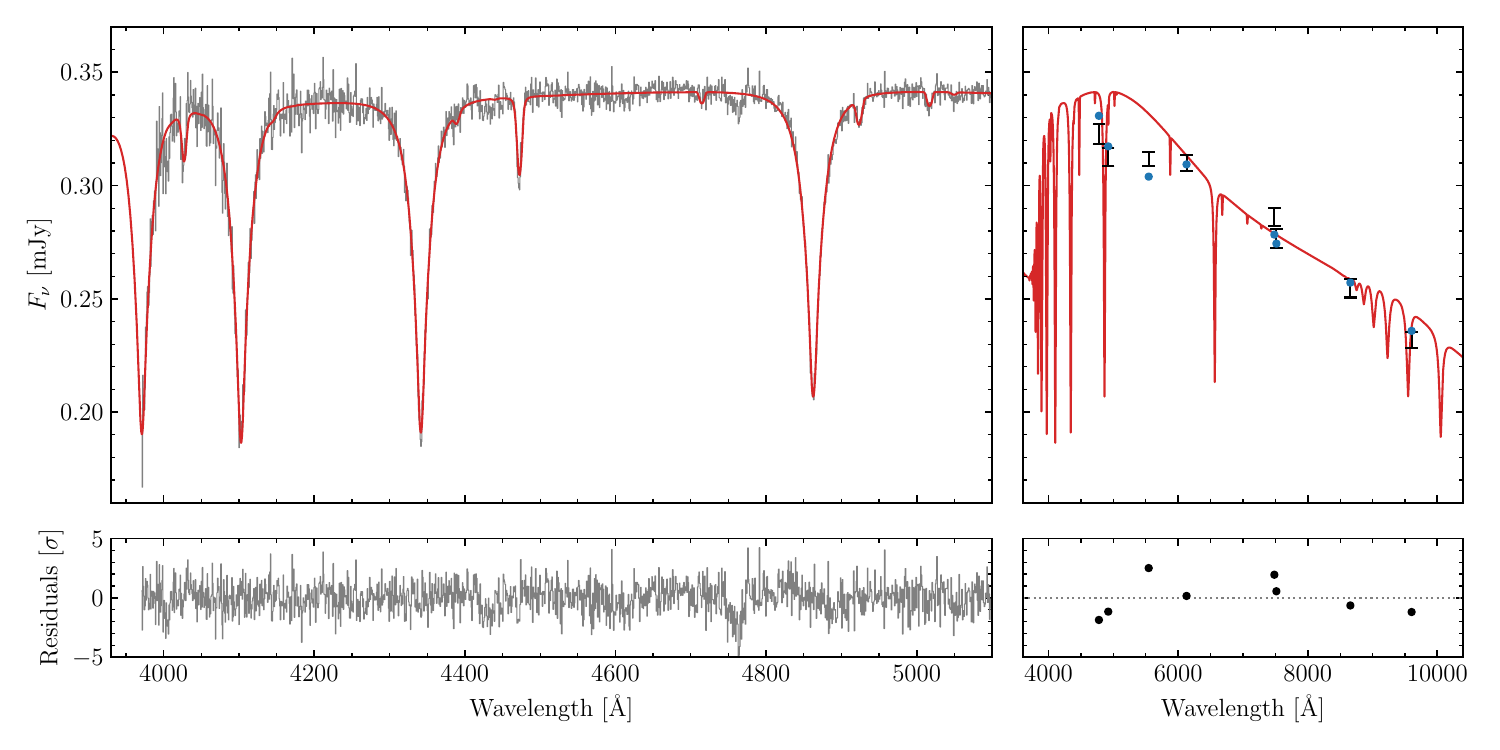}
   \caption{Fit to the  spectrum of J0526+5934 (left  panels) with the
     residuals shown below. The best  fitting model (red) includes the
     measured interstellar  reddening. The  spectral fluxes  have been
     re-calibrated against the model with a spline fit to their ratio.
     While the  GTC spectrum extends  to 5700\,\AA, it  is featureless
     beyond 5100\,\AA,  and so  those wavelengths  are not  shown. The
     same model  is shown against  the photometry (black  error bars),
     with  synthetic  photometry  shown  by  the  blue  points  (right
     panels).  Error bars  include  the  assumed 0.01\,mag  systematic
     uncertainty. Again, their residuals are shown below.}
\label{fig:spec_phot_fit}
\end{figure*}

\subsection{Radial velocities and radial velocity curve}
\label{s-rv}

In order to measure the radial velocities from the 24 GTC spectra (see
Section\,\ref{s-GTC}), we  first fitted the H$\beta$  absorption lines
with a  single Gaussian profile. We  used these values to  correct the
spectra from  the orbital motion,  which we  combined and run  a model
spectral  fit (see  details in  Section\,\ref{s-specfit}) to  obtain a
preliminary best-fit  model spectrum.  In a second  step, we  used the
\verb|MOLLY| software to cross-correlate the normalised best-fit model
spectrum  to  the  24  observed GTC  spectra  (also  normalised).  The
cross-correlation technique  yielded similar  but more  precise radial
velocity  values  than those  obtained  from  the Gaussian  fits.  The
observed spectra  were then  corrected from  the orbital  motion using
these refined velocities and the  combined spectrum was then re-fitted
to derive  the effective  temperature and surface  gravity of  the ELM
white dwarf (see details in Section\,\ref{s-specfit}).

Once the radial velocities were obtained using the above procedure, we
represented them as a function of orbital phase and fitted them with a
sine curve. The  radial velocity curve and the  corresponding sine fit
can be seen  in Figure\,\ref{f-rv_curve}. From this fit  we obtained a
semi-amplitude  velocity  of  565.2$\pm$3.2  km/s  for  the  brightest
component and a systemic velocity of $-21.6\pm2.2$ km/s.

\subsection{Spectral model fit and mass of the visible component}
\label{s-specfit}

To  measure the  atmospheric parameters  of  the ELM  white dwarf,  we
performed   a   simultaneous   fit   to  our   coadded   GTC   spectra
(Section\,\ref{s-GTC}) and the available photometry (\textit{Gaia} DR3
and PanSTARRS DR1). For this purpose, we used the Koester 1D LTE white
dwarf model atmosphere code \citep{Koester2010}, assuming a negligible
contribution     from    the     white     dwarf    companion     (see
Section\,\ref{s-teff2}). As  free parameters, we fitted  the effective
temperature ($T_\mathrm{eff}$),  the surface  gravity ($\log  g$), the
helium abundance ($\log(\mathrm{He}/\mathrm{H})$),  the solid angle of
the star on the sky ($\Omega$), the interstellar reddening ($E(B-V)$),
and finally  its rotational  broadening ($v_\mathrm{eq}\sin  i$; where
$v_\mathrm{eq}$ is the rotational velocity  at the equator and and $i$
is the inclination).

Our fit  consisted of an  iterative $\chi^2$ minimization  against the
spectrum and photometry. Rather than construct a grid of models around
the  approximate  solution  and  interpolating,  we  recalculated  our
atmosphere   code  at   each   iteration  in   the   fit,  to   ensure
self-consistency.  After computing  the  model at  the  start of  each
iteration,  we  convolved  the   model  spectrum  by  an  instrumental
broadening of  $R=2165$, as expected  for the R2000B grating.  We then
scaled the model  to observational fluxes by multiplying  by the solid
angle, $\Omega$. We  then applied the input  interstellar reddening to
the model,  and finally  applied rotational  broadening with  a kernel
determined from the Claret 4-term  limb darkening law evaluated in the
SDSS $g$-band.

We calculated synthetic  photometry in each of the  observed bands, by
integrating  the model  over  each bandpass,  to  compare against  the
observed fluxes,  where we assumed  each flux contained  an additional
0.01\,mag systematic  uncertainty. To compare with  the spectral data,
we normalised  the model against  the spectrum  using a spline  fit to
their spectral ratio (in order to remove the effects of imperfect flux
calibration).  We determined  the  total $\chi^2$  fit  by adding  the
individual $\chi^2$ for the spectrum  and photometry. We also included
the  measured reddening  $E(B-V)=0.27\pm0.05$ from  the 3D  extinction
maps of \citet{Lallement2014} as an additional data-point in the total
$\chi^2$  effectively  acting  as  a prior  on  our  fitted  reddening
parameter.

After      performing      this       minimization,      we      found
$T_\mathrm{eff}=27,330\pm370$\,K,                                $\log
g=6.213\pm0.050$\,dex(cm\,s$^{-2}$),
$\log(\mathrm{He}/\mathrm{H})=-2.20\pm0.03$\,dex,       $\Omega      =
(9.41\pm0.23)\times10^{-24}$\,sr,      $E(B-V)=0.383\pm0.007$,     and
$v_\mathrm{eq}\sin i=299\pm10$\,km\,s$^{-1}$. All quoted uncertainties
are determined from the covariance matrix  of the best fit. From these
we   determined   the  stellar   radius   to   be  $R_\mathrm{ELM}   =
0.065\pm0.005$\,\rsun, the ELM white  dwarf mass to be $M_\mathrm{ELM}
=    0.257\pm0.049$\msun,   and    their   Pearson    correlation   as
$\rho=0.785$. From the best fitting  model spectrum, we determined the
intrinsic  \textit{Gaia} absolute  magnitude to  be $G_\mathrm{abs}  =
6.73\pm0.17$\,mag  (with the  uncertainty  considering  the errors  on
$\Omega$ and the parallax) and its \textit{Gaia} colour to be $Bp-Rp =
-0.419\pm0.005$.

An  important source  of  uncertainty  in our  fit  is  the degree  of
rotational  broadening. Our  radius  and  orbital period  measurements
(assuming     tidal     locking)      suggest     $v_\mathrm{eq}     =
231\pm17$\,km\,s$^{-1}$, which  is notably  smaller than  our measured
value  of  $v_\mathrm{eq}\sin i$  (our  light  curve analysis  in  the
following  section  indicates  an inclination  of  $65\pm7$\,degrees).
While we cannot  provide a definite explanation  for this discrepancy,
we  acknowledge  that our  model  of  rotational broadening  does  not
account for  ellipsoidal distortion of  the star or  gravity darkening
which will certainly lead to more complex broadening. Nevertheless, we
do not believe  this invalidates our other stellar  parameters, as our
measured  $v_\mathrm{eq}\sin i$  simply  represents  the best  fitting
value with an incomplete model.  If we adopt an alternative broadening
profile such  a rectangular distribution  (which gives more  weight to
the  most  extreme   Doppler  shifts),  we  find  a   lower  value  of
$v_\mathrm{eq}\sin   i=247\pm10$\,km\,s$^{-1}$,    while   the   other
parameters are virtually unchanged.

\begin{figure}
   \centering
   \includegraphics[width=0.97\columnwidth]{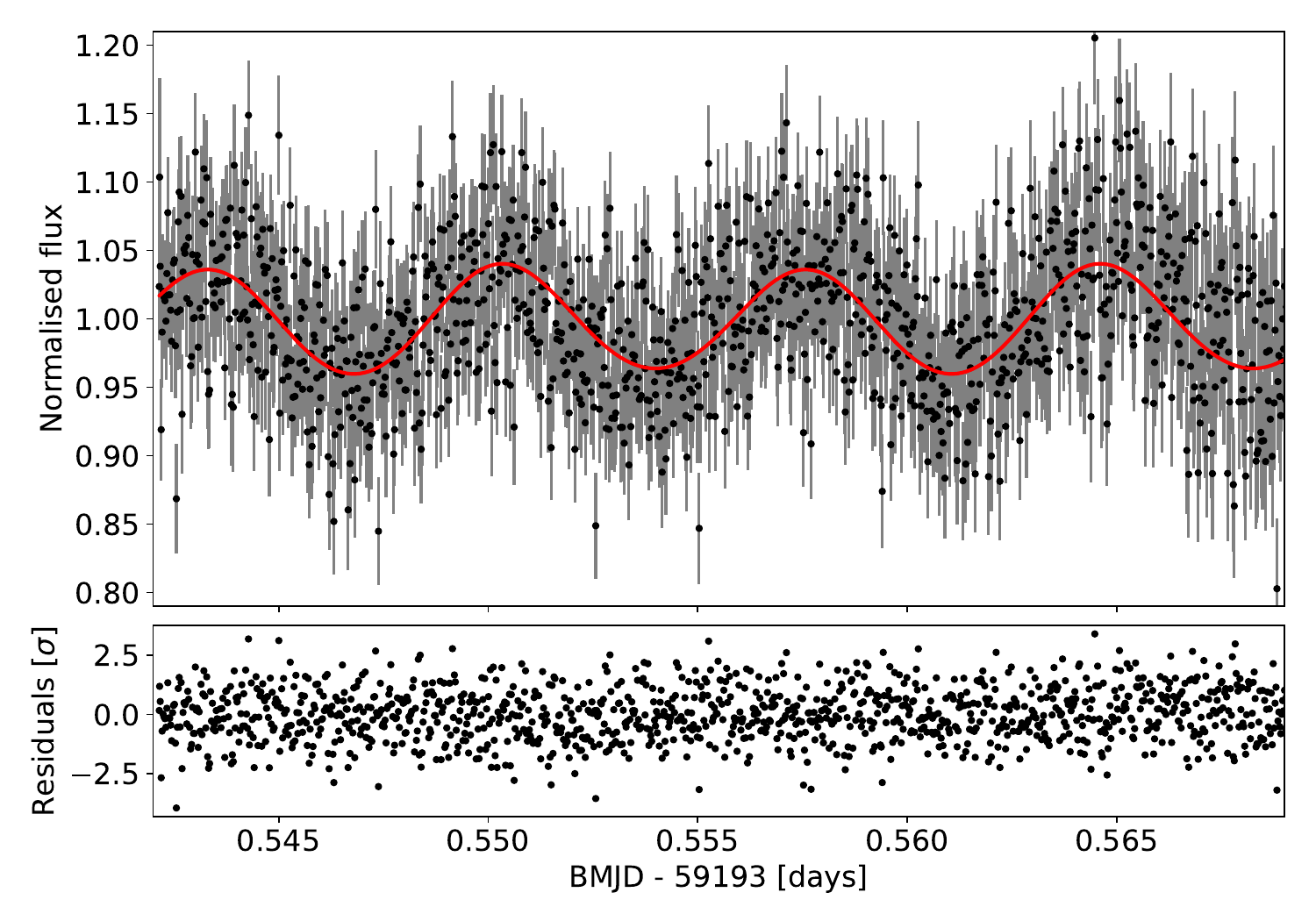}
   \caption{ULTRASPEC  $KG5$  band light  curve  with  best fit  model
     overplotted in red. The bottom  panel shows the residuals to this
     fit in terms of standard deviations from the model.}
\label{f-lc_fit}
\end{figure}

\subsection{Light curve fit}

The  ULTRASPEC light  curve shows  sinusoidal variations  on half  the
orbital period,  indicating that we are  seeing ellipsoidal modulation
originating from the tidally distorted ELM white dwarf. We can use the
amplitude  of this  variation to  set constraints  on the  stellar and
binary parameters since the fractional semi-amplitude is given by
\begin{equation}
\frac{\partial F}{F} = 0.15 \frac{(15 + u_1)(1+\beta_1)}{3 - u_1} \left( \frac{R_1}{a} \right)^3 q \sin^2 i, \label{e-ellip}
\end{equation}
\citep{Morris1993,Zucker2007} where $u_1$ is the linear limb-darkening
coefficient, $\beta_1$  is the gravity darkening  exponent, $R_1/a$ is
the radius scaled  by the orbital separation, $q=M_2/M_1$  is the mass
ratio and  $i$ is the inclination.  The subscript 1 refers  to the ELM
white dwarf, while 2 refers to the white dwarf companion.

In  the  absence  of  any  additional  information  when  fitting  the
ellipsoidal modulation  there is complete degeneracy  between the mass
ratio,    scaled    radius    of    the   ELM    white    dwarf    and
inclination. Therefore, we use the  constraints on the mass and radius
of  the ELM  white  dwarf (i.e.  $M_1$ and  $R_1$)  from the  spectral
modelling  to  break  some  of   this  degeneracy,  resulting  in  the
ellipsoidal amplitude depending only on $M_2$  and $i$ (since $a$ is a
function of  $M_1$, $M_2$ and P$_\mathrm{orb}$),  essentially allowing
us  to  place constraints  on  the  mass  of  the unseen  white  dwarf
companion from the light curve.

\begin{figure*}
   \centering
   \includegraphics[width=0.78\textwidth]{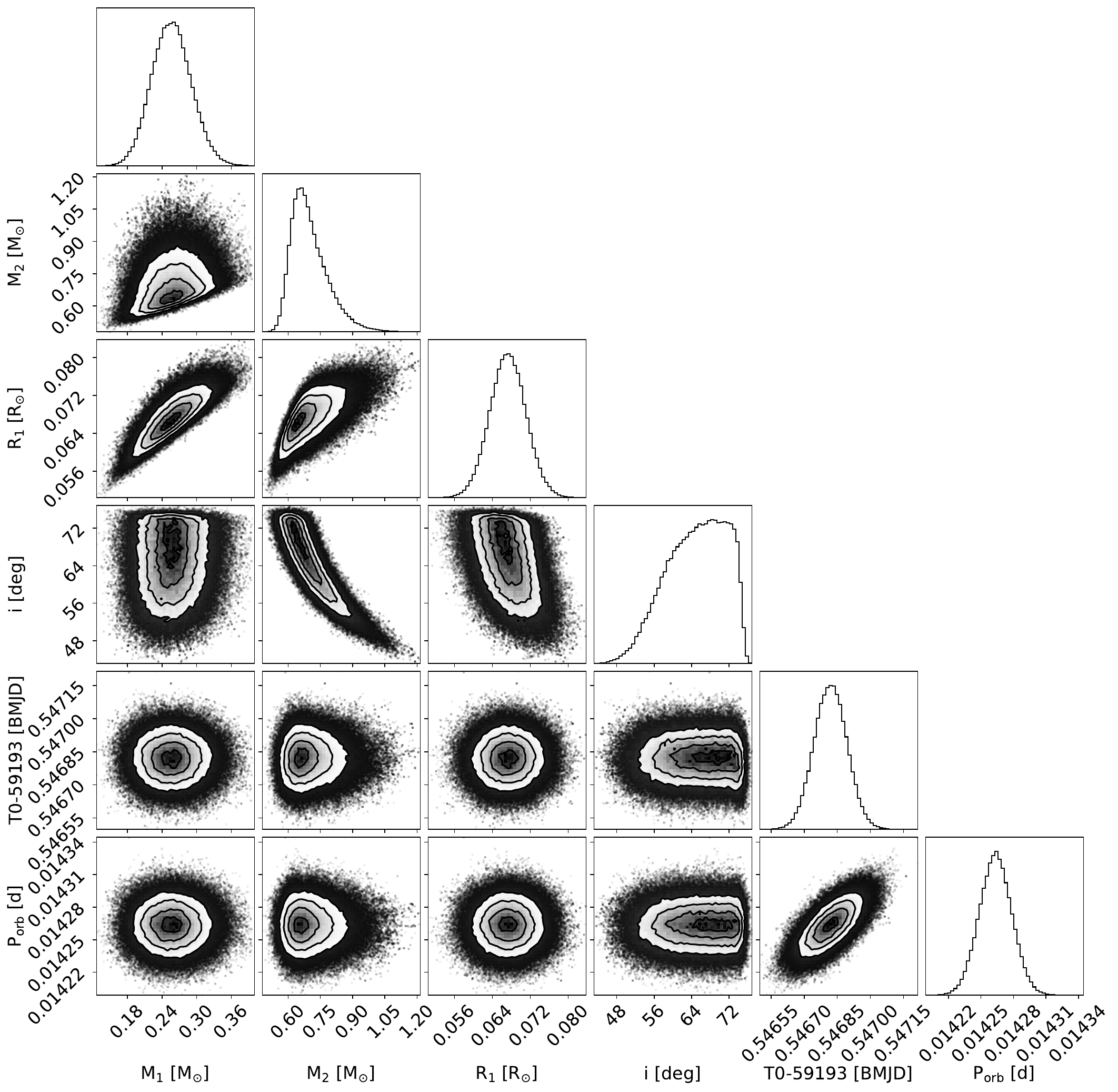}
   \caption{Posterior probability  distributions for  model parameters
     obtained  through   fitting  the   ULTRASPEC  $KG5$   band  light
     curve. Grey-scales  and contours represent the  joint probability
     distributions for  each pair of parameters,  while the histograms
     show   the  marginalised   probability  distributions   for  each
     parameter.}
\label{f-lc_params}
\end{figure*}

Rather  than simply  using Equation~\ref{e-ellip}  (which ignores  any
contribution from  the companion or  any irradiation effects  and does
not  allow  a more  sophisticated  treatment  of limb  darkening),  we
modelled the  light curve in  a more  complete way using  {\sc lcurve}
\citep{Copperwheat2010} and fitted the ULTRASPEC data using the Markov
chain Monte  Carlo (MCMC) method \citep{Press2007},  implemented using
the  Python  package  {\sc   emcee}  \citep{Foreman2013}.  The  fitted
parameters were: the mass of the  ELM white dwarf ($M_1$), the mass of
the white dwarf  companion ($M_2$), the radius of the  ELM white dwarf
($R_1$), the inclination  ($i$), the time of  the superior conjunction
of   the   ELM   white   dwarf  ($T_0$)   and   the   orbital   period
(P$_\mathrm{orb}$).   No   limb-darkening  coefficients   or   gravity
darkening exponents have been computed  for the parameter space of the
ELM white  dwarf in the  ULTRASPEC $KG5$ filter.  Therefore, following
the  method  in  \citet{Claret2020},  we   created  a  small  grid  of
limb-darkening (4-term law) and gravity  darkening values in the $KG5$
band for a white dwarf effective  temperature of 27,000\,K and a range
of surface  gravities between 6.0 and  6.5 in steps of  0.1 dex, using
updated model  DA spectra from  \citet{Koester2010}\footnote{Note that
  the dominant source of continuum opacity in the visible component of
  J0526+5934 is  from hydrogen. The additional  continuum opacity from
  the $\simeq$1\% of He in the atmosphere is negligible. Indeed, limb-
  and gravity-darkening  coefficients derived from a  pure-hydrogen DA
  model (i.e. without He traces) of the same effective temperature and
  surface gravity as our best-fit  model are nearly identical to those
  directly obtained from the best-fit model.}. Then for a chosen value
of  $M_1$ and  $R_1$ we  computed the  limb-darkening coefficient  and
gravity darkening  exponent of  the ELM  white dwarf  by interpolating
this grid.  We fixed  the temperature  of the ELM  white dwarf  to the
spectroscopically  determined value  of 27,330\,K  and also  fixed the
temperature of  the white  dwarf companion.  Given that  the companion
contributes a negligible  amount of flux to the  ULTRASPEC light curve
the  choice of  temperature  is somewhat  arbitrary. Nevertheless,  to
check that this parameter has no effect on the final results we fitted
the  light curve  twice, once  with the  temperature of  the companion
fixed at 8,000\,K and again with  it fixed at 20,000\,K. We also fixed
the  radius of  the  companion  to a  typical  white  dwarf radius  of
0.015\,{\rsun}. Again,  this parameter  makes little  difference given
the extreme flux ratio in the $KG5$ band.

We placed  a multivariate  Gaussian prior  on $M_1$  and $R_1$  with a
correlation   value   of   $\rho_{MR}   =   0.785$,   based   on   the
spectroscopically derived  values, which  takes into account  the fact
that  these  two  parameters  are strongly  correlated.  We  placed  a
Gaussian prior  on P$_\mathrm{orb}$ based  on the TJO and  GTC results
and a  uniform prior on the  inclination between 1 and  90 degrees. We
also used the radial velocity curve  to determine which minimum in the
light curve corresponded to the  superior conjunction of the ELM white
dwarf and  placed a uniform  prior on $T_0$  between the time  of this
minimum and $\pm 0.5$\,P$_\mathrm{orb}$ to ensure that the fit did not
try to jump  to the next cycle.  Finally, at each step  in the fitting
process  we computed  the radial  velocity semi-amplitude  of the  ELM
white   dwarf   (which  is   only   a   function  of   $M_1$,   $M_2$,
P$_\mathrm{orb}$ and $i$) and compared  this to the measured value via
a Gaussian  prior based on  the measured value. This  final constraint
helps break some of the degeneracy between $M_2$ and $i$.

Our MCMC fitting  used 50 walkers, each with 20,000  points. The first
5,000 points were classed as "burn-in" and were removed from the final
results. The  ULTRASPEC light  curve and  best fit  model is  shown in
Figure~\ref{f-lc_fit}, while  the posterior  probability distributions
are  shown  in  Figure~\ref{f-lc_params}.  As  expected  we  found  no
difference  in   the  results   between  an  8,000\,K   and  20,000\,K
companion. We found  that inclinations greater than  around 75 degrees
are  excluded, since  we  would  expect to  see  the  ELM white  dwarf
eclipsed by  its companion  at these high  inclinations (which  is not
seen in  the light curve).  Inclinations lower than around  40 degrees
are also ruled out, since at these low inclinations it is not possible
to generate a large enough ellipsoidal signal in the light curve while
simultaneously  being consistent  with  the  measured radial  velocity
semi-amplitude of the  ELM white dwarf. The best  consistency is found
at higher inclinations,  hence lower companion white  dwarf masses, as
reflected in  the marginalised  probability distribution for  $M_2$ in
Figure~\ref{f-lc_params}.  Overall  we  constrain the  inclination  to
$65\pm7$   degrees   and   the   companion   white   dwarf   mass   to
$0.71^{+0.09}_{-0.06}$\,\msun.  This  mass   is  consistent  with  our
assumption  that   the  unseen  companion   is  a  white   dwarf.  The
distributions  for  $P_\mathrm{orb}$,  $M_1$ and  $R_1$  are  entirely
driven by the priors placed on these values, while we find a value for
$T_0$  of  BMJD(TDB)  =  59193.54682(7),   where  the  number  in  the
parenthesis represents the uncertainty on the final digit.

\begin{table*}
\centering
\caption[]{Orbital and  stellar parameters  for the  visible component
  and  the  unseen companion  of  J0526+5934.  We include  the  values
  measured   by   \citet{Kosakowski2023}    and   \cite{Lin2023}   for
  comparison.} \label{t-param}
\begin{tabular}{lccc}
\hline
\hline
Binary & This work & \citet{Kosakowski2023} & \citet{Lin2023} \\
\hline
Orbital period (s)  & 1232.00$\pm$  0.66 & 1230.37467$\pm$0.00007& 1230.374556$\pm$0.000318\\
Orbital inclination (º)  & 65 $\pm$ 7 & 57.1$^{+4.3}_{-4.1}$ & 68.2$^{+3.7}_{-5.2}$\\
Systemic velocity (km/s) & -21.6 $\pm$ 2.2 & -40.7 $\pm$ 4.1& -35.6 $\pm$ 4.4\\
\hline
 Visible component & &  & \\
 \hline
T$_\mathrm{eff}$ (K) & 27,330 $\pm$ 370 & 27,300 $\pm$ 260& 25,480 $\pm$ 360\\
log\,g (dex)  & 6.213  $\pm$0.050 & 6.37  $\pm$0.03& 6.355  $\pm$0.068\\
M (M$_{\odot}$)   & 0.257 $\pm$ 0.049 & 0.378$^{+0.066}_{-0.060}$ & 0.360$^{+0.080}_{-0.071}$ \\
R (R$_{\odot}$)   & 0.065 $\pm$ 0.005 & 0.070 $\pm$ 0.005& 0.0661 $\pm$ 0.0054\\
$[$He/H$]$ (dex) & -2.20 $\pm$ 0.03 & -2.45 $\pm$ 0.06 & -2.305 $\pm$ 0.062 \\
K (km/s)  & 565.2 $\pm$  3.2 &558.3 $\pm$ 4.8 & 559.6$^{+6.4}_{-6.5}$\\
\hline
Unseen WD & &   &  \\
\hline
M (M$_{\odot}$)   & $0.71^{+0.09}_{-0.06}$ & $0.887^{+0.110}_{-0.098}$& $0.735^{+0.075}_{-0.069}$\\
T$_\mathrm{eff}$ (K)  & $<$6700 &- &- \\
\hline
\end{tabular}
\end{table*}

\subsection{Effective temperature of the white dwarf companion}
\label{s-teff2}

We attempted to  obtain an upper limit to the  white dwarf companion's
effective  temperature  by  determining  by  how  much  the  synthetic
spectrum of a (hydrogen-rich) white dwarf can be added to the spectrum
of  the  ELM  white  dwarf without  affecting  the  observed  spectral
features. To that end we considered the best-fit model to the observed
GTC  combined spectrum  from  Section\,\ref{s-specfit} and  subtracted
model spectra  from the \citet{Koester2010} updated  collection of any
given effective  temperature and surface  gravity between 7.8  and 8.3
dex  (that  is  a  surface  gravity that  corresponds  to  a  mass  of
0.71$^{+0.09}_{-0.06}$\,M$_\mathrm{\odot}$, as derived  from the light
curve fit). After subtracting each  white dwarf model, we measured the
resulting  H$\beta$ equivalent  width  (EW). We  assumed the  spectral
features to be  different from those sampled by the  best-fit model to
the observed spectrum when the EW decreased by more than 25\%. Fitting
such  spectra  would  result  in  stellar  parameters  that  would  be
different  to those  obtained in  Section\,\ref{s-specfit} due  to the
change in  the Balmer  line profiles.  However, we  found that  the EW
remain unaltered in all cases, with  the ELM contribution to the total
flux being always larger than 96\%.  This implied no constraint on the
unseen white  dwarf companion's effective temperature  could be placed
with  this exercise.  More  stringent constraints  on  this value  are
discussed in Section\,\ref{s-discuss}.

\subsection{Comparison to \citet{Kosakowski2023} and \citet{Lin2023}}

The  most relevant  orbital and  stellar parameters  we have  obtained
throughout   this    section   for   J0526+5934   are    provided   in
Table\,\ref{t-param}.  All our  measured values  are similar  to those
recently  obtained  by   \citet{Kosakowski2023}  and  \citet{Lin2023},
although slight differences arise (see Table\,\ref{t-param}). The most
important difference in the context of  this paper is that we derive a
lower surface gravity for the visible component and, as a consequence,
a lower mass. In the three studies, these values are directly obtained
from the spectral fitting  analysis. Whilst our observational strategy
ensured    we    are    not   suffering    from    orbital    smearing
(Section\,\ref{s-rv}),   the   exposure    times   used   during   the
spectroscopic observations performed by \citet{Kosakowski2023} covered
$\simeq$10\%  of  the orbit,  therefore  their  combined spectrum  was
considerably  affected by  smearing. The  observations carried  out by
\citet{Lin2023} were clearly affected by this effect too. Although the
authors fitted all  their spectra simultaneously to  account for this,
our results should  yield more accurate values  simply because orbital
smearing  is efficiently  taken into  account. As  a consequence,  our
derived mass  provides robust  evidence for  the visible  component in
J0526+5934 not  to necessarily be a  hot subdwarf star, but  rather an
ELM  white dwarf,  as we  have  been assuming  in this  paper. In  the
following section we discuss this hypothesis in detail.

\section{Discussion}
\label{s-discuss}

The deduced stellar mass values for the bright component of J0526+5934
strongly reduce  the likelihood of  it being  a hot subdwarf  star. In
fact,  the minimum  mass  requirement for  a  subdwarf star  typically
stands  at approximately  0.30\,M$_{\odot}$ \citep{Arancibia2024}.  In
particular,  detailed  calculations  by \citet{Han2003}  predict  this
minimum mass  to be about  0.33 M$_{\odot}$. This  is due to  the fact
that  stars  below   this  mass  threshold  are   unable  to  initiate
core-helium   burning   within   their   cores   under   nondegenerate
conditions.   From  the   observational  point   of  view,   the  mass
distribution of  hot subdwarf  stars has been  found to  range between
$\simeq$0.3\,M$_{\odot}$ and  0.63\,M$_{\odot}$, with a clear  peak at
$\simeq$0.45\,M$_{\odot}$     and    very     few    objects     below
$\simeq$0.3\,M$_{\odot}$     \citep{Fontaine2012,    Schaffenroth2022,
  Lei2023}.  This gives  further support  to our  hypothesis that  the
visible component of J0526+5934 is an ELM white dwarf.

ELM white  dwarfs are  expected to originate  from unstable  mass loss
through the common envelope ejection channel and the stable Roche lobe
overflow  channel, as  discussed recently  by \cite{Li2019}.  Existing
evolutionary tracks for  ELM white dwarfs are based  on the assumption
of stable  Roche lobe  overflow, involving  stable mass  transfer (see
\citealt{Althaus2013,Istrate2016}            and            references
therein). Consequently,  ELM white  dwarf evolutionary  models derived
under  this framework  are  characterized  by an  upper  limit on  the
possible H-layer thickness that a given-mass ELM white dwarf model can
possess. This implies that residual stable hydrogen burning emerges as
the primary energy source during the cooling phase of the white dwarf,
leading to the occurrence of multiple hydrogen flashes at the onset of
the cooling track.

The  manifestation of  hydrogen flashes  results in  diverse potential
solutions for  the observed  ELM white dwarf  component. Specifically,
from  the evolutionary  sequences computed  by \cite{Althaus2013},  we
deduced a  stellar mass  of 0.237$\pm$0.035\,M$_{\odot}$, a  radius of
0.06$\pm$0.006\,R$_{\odot}$, a cooling age  of 260$\pm$240\,Myr, and a
helium surface  abundance [He/H]  in the  range of  -2. and  -4. These
values  perfectly  agree with  the  observational  inferences for  the
J0526+5934  bright component  (Section\,\ref{s-specfit}), thus  giving
strong support  for the visible  object to be  an ELM white  dwarf. In
particular,  the  derived   mass  value  is  in   agreement  with  the
predictions  of population  property studies  of ELM  white dwarfs  in
double degenerate  systems, which  suggests that intrinsic  masses for
ELM  white  dwarfs peak  around  0.25M$_{\odot}$  for the  CE  channel
\citep{Li2019}. The considerable range in derived cooling ages are the
result of residual hydrogen burning in the envelope.

It  is  noteworthy that  the  helium  abundance predicted  by  cooling
sequences is  also in  agreement with  our independent  value obtained
from       the       spectral        analysis       performed       in
Section\,\ref{s-specfit}.  Gravitational   settling  rapidly  depletes
metals in the  atmospheres of white dwarfs; however, due  to the lower
surface  gravities  characterizing ELM  white  dwarfs,  the impact  of
gravitational settling is less  pronounced.  This explains that traces
of helium persist in the envelope  at the effective temperature of the
ELM component. For low-mass ELM white dwarfs, a pure hydrogen envelope
is expected, albeit at significantly lower effective temperatures. The
helium   abundances   predicted   by    the   cooling   sequences   of
\cite{Althaus2013}  are  likely  a conservative  lower  limit.  Higher
helium  abundances could  potentially  be expected  due to  rotational
mixing,  countering  the  effects  of gravitational  settling  in  the
surface layers  of young ELM  proto-white dwarfs. However, as  the ELM
contracts  and  embarks  on  its  cooling  track,  the  efficiency  of
rotational mixing diminishes,  and the role of  rotation in augmenting
helium   abundance  becomes   less   significant,   as  discussed   by
\cite{Istrate2016}.

The formation  of ELM white  dwarfs with hydrogen contents  lower than
expected   from   stable  Roche   lobe   overflow   cannot  be   ruled
out. Specifically, \cite{Strickler2009} concluded that the presence of
a  population of  low-mass  He-core white  dwarfs  with thin  hydrogen
envelopes   in  NGC   6397   cannot  be   discarded.  More   recently,
\cite{Irrgang2021} demonstrated the necessity of  a thin H envelope in
J1604+1000,  a  proto-ELM  white  dwarf  weighing  approximately  0.21
\msun\, likely  resulting from  a common  envelope event.  Moreover, a
comparative analysis  of the mass  distribution of observed  ELM white
dwarfs with  theoretical expectations  by \cite{Li2019}  has suggested
that the  ELM white  dwarfs arising from  the common  envelope channel
could  be  characterized  by  thinner H  envelope  compared  to  those
resulting from stable Roche lobe channel. Finally, asteroseismological
analysis  support the  presence  of  thin H  envelopes  (1-2 order  of
magnitude thinner  than predicted  by the  stable Roche  lobe overflow
channel) in certain pulsating ELM white dwarfs \citep{Leila2018}.

While the majority of ELM white  dwarfs are discovered in double white
dwarf  systems resulting  from  common envelope  episodes, nearly  all
evolutionary  calculations  that  involve  stable  mass  transfer  are
applied to derive the properties of the white dwarf components in such
systems.  However, it  could  be  expected that  in  ELM white  dwarfs
resulting from  common envelope, the recurrent  hydrogen shell flashes
may not  occur, because of the  minor role of residual  H burning.  In
fact, the evolution  of proto-ELM white dwarfs depends on  the mass of
the hydrogen-rich layer above the  helium core, which is determined by
the  detailed common  envelope  ejection process,  the most  uncertain
phase in binary evolution \citep{Li2019}. A reduction in the thickness
of the  hydrogen envelope by  a factor  of 2 causes  residual hydrogen
burning  to   become  negligible  \citep{Leila2018thin}.   With  these
considerations  in mind,  i.e. neglecting  the occurrence  of hydrogen
flashes,   we  should   expect   a  stellar   mass  of   approximately
0.27M$_{\odot}$,   also   in    agreement   with   the   observational
inference. The resulting cooling times are expected to be much shorter
than  the  cooling times  of  their  counterpart with  thick  hydrogen
envelope (see \citealt{Leila2018thin}).

From the  observational data of  J0526+5934 and based on  the probable
initial configuration of the system and cooling times, we can infer an
upper limit  on the  effective temperature of  the unseen  white dwarf
companion. According to \citet{Li2019}, the most likely progenitors of
ELM white dwarfs from common envelope channel have masses in the range
0.95-1.25\,M$_{\odot}$.  Given that  the  common  envelope channel  is
responsible for ELM  white dwarfs in double degenerate  systems with a
helium-core  white  dwarf  more  massive  than  about  0.22M$_{\odot}$
\citep{Li2019}, we  adopt a  value of  1.25M$_{\odot}$ as  the maximum
stellar mass for the progenitor of the ELM component. This would yield
an age of  approximately 3.8 Gyr for the progenitor  to leave the main
sequence \citep{Miller2016}. Considering the  potential mass range for
the unseen  white dwarf companion (Table\,\ref{t-param}),  the minimum
mass  of its  progenitor would  be on  the order  of 3.0\,M$_{\odot}$,
suggesting    a     main    sequence     lifetime    of     0.3    Gyr
\citep{Miller2016}.  Thus, the  newly formed  white dwarf  should have
been undergoing cooling for at least  3.5 Gyr. During this period, and
depending on its mass, the invisible white dwarf component should have
attained   an    effective   temperature    within   the    range   of
5700-6700K. These  inferred values  serve as upper  limits considering
that the  ELM could have  arisen from  a progenitor less  massive than
1.25M$_{\odot}$.

By considering  J0526+5934 as  a compact binary  star composed  of two
white  dwarfs of  masses  of  0.26\,M$_{\odot}$ and  0.71\,M$_{\odot}$
(Table\,\ref{t-param}),  we  estimate  the  merger  time  due  to  the
emission of  gravitational waves ($\tau$;  in Myr) from  the following
expression:
\begin{equation}
\tau = 47925 \frac{(M_1 + M_2)^{1/3}}{M_1 M_2} P^{8/3},  \\
\end{equation}
where $P$ is the orbital period in days and the masses are in units of
M$_{\odot}$     \citep{Kraft1962}.     This    results     in     just
$\simeq$3\,Myr. The result of the  merger, assuming no mass loss takes
place     during    the     process,     would     be    a     massive
($\simeq$1\,M\,$_{\odot}$) white dwarf.  A relatively large percentage
of such  massive white dwarfs is  expected to arise from  this kind of
mergers \citep{Cheng2019, Temmink2020, Kilic2023}, which can partially
explain  the  high-mass  excess  observed  in  the  white  dwarf  mass
distributions over  the last years  \citep{Rebassa2015b, Rebassa2015a,
  Jimenez2023, Kilic2020}. It is also possible that the system ends up
as  a  .Ia  supernovae.  As indicated  by  \citet{Shen2015},  in  this
scenario  the  ELM  white  dwarf  would  begin  transferring  material
(sufficiently enriched in  He) and a He detonation  would occur during
the subsequent  merger of the two  white dwarfs, thus producing  a .Ia
supernovae.

\section{Summary and conclusions}
\label{conclusions}

We have  presented an independent observational  and theoretical study
of  the ultra-short  period binary  J0526+5934. The  observations were
carried out at the Telescopi  Joan Or\'o, The Gran Telescopio Canarias
and the Thai National Telescope,  and the evolutionary models employed
were those  from the La Plata  group. We have provided  constraints on
the orbital  and stellar  parameters of  both stellar  components. Our
results  are similar  to those  obtained by  the previous  analysis of
\citet{Kosakowski2023} and  \citet{Lin2023}. However,  we find  a mass
for  the visible  component  of  0.26$\pm$0.05\,M$_{\odot}$ (from  the
observational  analysis)  or  0.237$\pm$0.035\,M$_{\odot}$  (from  the
theoretical  evolutionary tracks).  This  is lower  than the  reported
values       of        $0.378^{+0.066}_{-0.060}$\,M$_{\odot}$       by
\citet{Kosakowski2023}  and $0.360^{+0.080}_{-0.071}$\,M$_{\odot}$  by
\citet{Lin2023}. This  difference is key when  interpreting the nature
of the visible component. Whilst the results of \citet{Kosakowski2023}
and \citet{Lin2023} favour  a hot subdwarf star, our  analysis is much
more in  agreement with an  extremely low-mass white dwarf.  The three
studies cannot completely rule out the alternative scenario and future
observations may shed further insight on this remarkable object.

\begin{acknowledgements}

This  work  was  partially  supported  by  the  Spanish  MINECO  grant
PID2020-117252GB-I00 and  by the AGAUR/Generalitat de  Catalunya grant
SGR-386/2021. SGP acknowledges the support of a Science and Technology
Facilities   Council   (STFC)   Ernest   Rutherford   Fellowship.   IP
acknowledges  support   from  a  Royal  Society   University  Research
Fellowship    (URF\textbackslash    R1\textbackslash    231496).    RR
acknowledges support  from Grant RYC2021-030837-I funded  by MCIN/AEI/
10.13039/501100011033   and   by    “European   Union   NextGeneration
EU/PRTR”.   M.C.  acknowledges   grant  RYC2021-032721-I,   funded  by
MCIN/AEI/10.13039/501100011033   and    by   the    European   Union46
NextGenerationEU/PRTR.  This  research  made  use  of  data  from  the
European     Space     Agency     (ESA)     mission     {\it     Gaia}
(\url{https://www.cosmos.esa.int/gaia}), processed  by the  {\it Gaia}
Data      Processing      and     Analysis      Consortium      (DPAC,
\url{https://www.cosmos.esa.int/web/gaia/dpac/consortium}).    Funding
for the DPAC has been provided by national institutions, in particular
the  institutions   participating  in  the  {\it   Gaia}  Multilateral
Agreement.

This work makes use of data from the 80 cm Telescopi Joan Oró (TJO) of
the Montsec Astronomical Observatory  (OAdM), owned by the Generalitat
de  Catalunya and  operated  by the  Institut  d’Estudis Espacials  de
Catalunya  (IEEC).  Based on  observations  obtained  with the  Samuel
Oschin Telescope  48 inch  and the  60 inch  telescope at  the Palomar
Observatory as part of the Zwicky Transient Facility project. Based on
observations  made  with  the   Gran  Telescopio  Canarias  (programme
GTC6-21B),  installed in  the Spanish  Observatorio del  Roque de  los
Muchachos of the  Instituto de Astrofísica de Canarias,  in the island
of La Palma.
 
\end{acknowledgements}

\bibliographystyle{aa}
\bibliography{DD}



\end{document}